\documentclass[twocolumn,showpacs,preprintnumbers,superscriptaddress]{revtex4}
\usepackage{amsmath}
\usepackage{dcolumn}
\usepackage{bm}
\usepackage{epsfig}
\usepackage{graphicx}
\usepackage{amsfonts}
\usepackage{amssymb}
\usepackage{mathrsfs}
\usepackage{color}
\usepackage{appendix}

\begin{document}

\title{A complicated Duffing oscillator in the surface-electrode ion trap}
\author{Hao-Yu Wu}
\affiliation{State Key Laboratory of Magnetic Resonance and Atomic and Molecular Physics,
Wuhan Institute of Physics and Mathematics, Chinese Academy of Sciences,
Wuhan 430071, China}
\affiliation{University of the Chinese Academy of Sciences, Beijing 100049, China}
\author{Yi Xie}
\affiliation{State Key Laboratory of Magnetic Resonance and Atomic and Molecular Physics,
Wuhan Institute of Physics and Mathematics, Chinese Academy of Sciences,
Wuhan 430071, China}
\affiliation{University of the Chinese Academy of Sciences, Beijing 100049, China}
\author{Wei Wan}
\affiliation{State Key Laboratory of Magnetic Resonance and Atomic and Molecular Physics,
Wuhan Institute of Physics and Mathematics, Chinese Academy of Sciences,
Wuhan 430071, China}
\affiliation{University of the Chinese Academy of Sciences, Beijing 100049, China}
\author{Liang Chen}
\affiliation{State Key Laboratory of Magnetic Resonance and Atomic and Molecular Physics,
Wuhan Institute of Physics and Mathematics, Chinese Academy of Sciences,
Wuhan 430071, China}
\author{Fei Zhou}
\affiliation{State Key Laboratory of Magnetic Resonance and Atomic and Molecular Physics,
Wuhan Institute of Physics and Mathematics, Chinese Academy of Sciences,
Wuhan 430071, China}
\author{Mang Feng}
\email[Corresponding author Email: ]{mangfeng@wipm.ac.cn}
\affiliation{State Key Laboratory of Magnetic Resonance and Atomic and Molecular Physics,
Wuhan Institute of Physics and Mathematics, Chinese Academy of Sciences,
Wuhan 430071, China}

\begin{abstract}
The oscillation coupling and different nonlinear effects are observed in a
single trapped $^{40}Ca^{+}$ ion confined in our home-built
surface-electrode trap (SET). The coupling and the nonlinearity are
originated from the high-order multipole potentials due to different layouts
and the fabrication asymmetry of the SET. We solve a complicated Duffing
equation with coupled oscillation terms by the multiple scale method, which
fits the experimental values very well. Our investigation in the SET helps
for exploring nonlinearity using currently available techniques and for
suppressing instability of qubits in quantum information processing with
trapped ions.
\end{abstract}

\date{\today }
\maketitle

\section{introduction}

The Duffing oscillator is generally used to describe nonlinear dynamics in
oscillating systems \cite{Nayfeh1979,M.I. Dykman,L. D. Landau,Nayfeh1981}.
The corresponding Duffing equation models a damping and driven oscillator
with more complicated behavior than simple harmonic motion, which can be
used to exhibit chaos in dynamics and hysteresis in resonance \cite{V. I.
Arnold,S. H. Strogatz,H. B. Chan,M. I. Dylman2004,B. Yurke,E. Buks}.

On the other hand, the motion of trapped ions is highly controllable and can
be employed to transfer quantum information when cooled down to ground state
\cite{blatt}. Since it is effectively approximated to be harmonic, the ion
motion in a quadruple electromagnetic trap \cite{W. Paul} can be regarded as
a good mechanical oscillator, which may exhibit nonlinearity when driven to
the nonlinear field. For example, Duffing nonlinear dynamics has been
investigated in a single ion confined in the linear ion trap \cite{aker}.
The trap nonlinearity introduces instability in the motion of the ion, which
should be avoided in most times, but can also be used in resonance rejection
and parameter detection in mass spectrometry \cite{mak, dra}. Recent
research also showed the feasibility of phonon lasers based on the
nonlinearity of a single trapped ion under laser irradiation \cite{vahala}.

We focus in this work on the nonlinearity in a home-built surface-electrode
trap (SET). The SET, with capability to localize and transport trapped ions
in different potential wells, is a promising setup for large-scale quantum
information processing \cite{wineland}. In comparison to conventional linear
Paul traps, however, the reduced size and asymmetry in SET lead to stronger
high-order multipole potentials \cite{wesen,house,Blakestad}, which affect
the stability of the ion trapping. To solve the problem we have to
understand the source and the strength of the nonlinearity. Due to
complexity resulted from the high-order multipole potentials, the nonlinear
effect in the SET cannot be simply described by the Duffing oscillator as
for the linear trap, but an inhomogeneous-coupled Duffing oscillator
involving quadratic and cubic nonlinearities. We observed the nonlinearity experimentally
in our SET, and by the method of multiple scales we derived an
inhomogeneous-coupled Duffing oscillator to fit the experimental values,
which shows that both the nonlinearity and axial-radial coupling exist in
the case of the frequency resonance (i.e., around the regime of driving
detuning being zero). Moreover, we show in the non-coupling case different
nonlinear effects in different dimensions, which is due to different
asymmetry in fabrication of the SET.

\section{experimental setup and images of ion motion}

\begin{figure}[tbph]
\includegraphics[width=8.5 cm,bb=88pt 459pt 511pt 732pt]{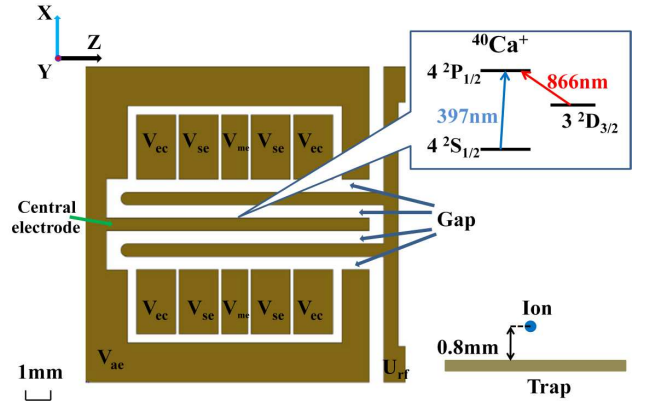}
\caption{The layout of our home-built SET with the top-right inset for the
relevant energy levels of the trapped $^{40}Ca^{+}$ ion. The $^{40}Ca^{+}$
ion is cooled by the laser beam at 397 nm on the $4s^{2}S_{1/2}%
\longleftrightarrow 4p^{2}P_{1/2}$ transition, helped by another laser light
at 866 nm as repumping. Here, the trapped ion is above the electrodes by 0.8
mm (See the bottom-right inset).}
\end{figure}

Our home-built SET is a 500 $\mu $m-scaled planar trap with five electrodes
for radial confinement, and fabricated by printed circuit board technology
\cite{chip}. As shown in Fig. 1, the five electrodes consists of a central
electrode, two radio-frequency (rf) electrodes and two outer segmented dc
electrodes, where the rf electrodes, the central electrode and the gaps in
between are of the same width of 500 $\mu$m. Each outer segmented electrode
consists of five component electrodes, i.e., a middle electrode, two control
electrodes and two end electrodes. The widths of the control electrodes and
end electrodes in the segment are 1.5 mm and the middle electrode is 1 mm
wide. The gap in the segmented electrode is of 500 $\mu$m width. When the
SET works, the trapped $^{40}Ca^{+}$ ion stays above the electrodes by $0.8$
mm, and the pseudopotential trapping depth is below $1$ eV with rf amplitude
$U$(0-peak)$=400$ V and rf frequency $\Omega /2\pi =15$ MHz. The voltage on
the four end electrodes is $V_{ec}=40$ V but zero on other electrodes.

\begin{figure}[tbph]
\includegraphics[width=8.5 cm,bb=83pt 474pt 504pt 770pt]{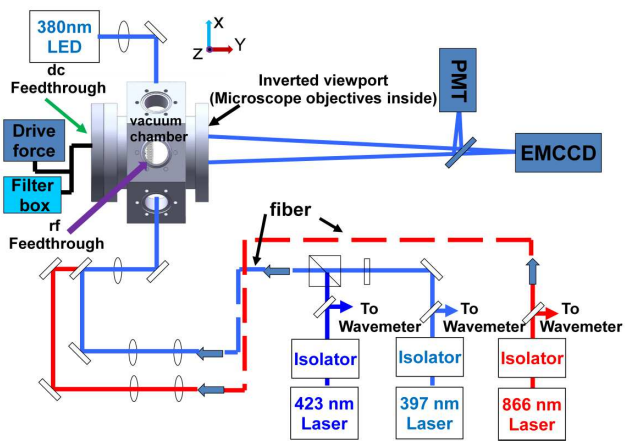}
\caption{Schematic of our experimental setup. The main components are
explained in the text.}
\end{figure}

The experimental setup is plotted in Fig. 2, where the ultraviolet radiation
at $397$ nm excites $4s^{2}S_{1/2}\longleftrightarrow 4p^{2}P_{1/2}$
transition by a grating stabilized laser diode with power up to $30$ mW and
linewidth less than 2 MHz. Another grating stabilized laser diode at $866$
nm with power up to 100 mW and linewidth less than 5 MHz excites the $%
3d^{2}D_{3/2}\longleftrightarrow 4p^{2}P_{1/2}$ transition. The frequencies
of the both laser diodes have been calibrated to the wavelength meter
(HighFiness, WS-7). Typical laser powers at the trap center are 50 $\mu$W
for 397 nm in red detuning (-80 MHz) and 500 $\mu$W for 866 nm in carrier
transition. In our SET, the single $^{40}Ca^{+}$ ion is laser cooled and
stably confined, which is monitored by photon scattering collected by both
an electron-multiplying charge-coupled-device (EMCCD) camera and a
photomultiplier tube (PMT). Outside the vacuum chamber, the electrical
connections immediately encounter a "filter box", which provides low-pass
filtering of the voltages applied to the electrode. An additional drive
force is electrically connected to one of the middle electrodes behind the
filter box, which provides an excitation to drive the ion away from
equilibrium. Due to the design of our SET system, the motion of the ion is
detected only in the $xz$ plane by the EMCCD. As a result, what we study
throughout the work is the oscillation along the axial direction ($z$-axis)
and the radial direction ($x$-axis), whose harmonic frequencies are,
respectively, $\omega _{0z}/2\pi =191.7$ kHz and $\omega _{0x}/2\pi =425$
kHz. Moreover, since the harmonic frequency in $y$-axis is $\omega
_{0y}/2\pi =925$ kHz, much bigger than in other axes, the ion can be
regarded as a very tight confinement in y direction. We have suppressed the
micro-motion by the rf-photon cross correlation compensation \cite{njp},
which yield cooling of the ion down to the temperature below 10 mK.

To study the nonlinear mechanical response, we drive the ion to the
nonlinear regime by a small oscillating voltage, i.e., $V=7$ V, applied to
one of the middle electrodes. We slowly increase the driving frequency with
the scan step $0.1$ kHz, from $189.0$ kHz to $433.0$ kHz (the positive
scan), the ion oscillates first along the $z$-axis and then turns to the $x$%
-axis for oscillation. The particularly interesting observation is the
simultaneous responses, i.e., a rectangle trajectory, in both $x$- and $z$-
axes when the sweep is close to the harmonic resonator frequency in either
of the axes. Similar behavior is also found in the negative scan.
\begin{figure}[tbph]
\includegraphics[width=8.5 cm,bb=84pt 482pt 506pt 772pt]{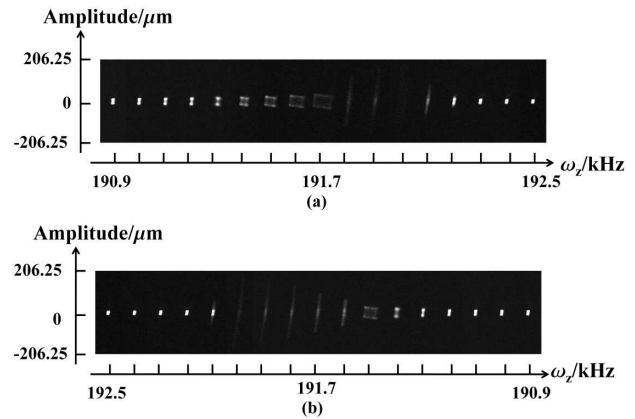}
\caption{Time-averaged images of a single trapped ion taken at different
drive frequencies, with (a) the positive scan and (b) the negative scan. The
prime oscillation is along the axial direction and the scan step is 0.1 kHz.
The rectangles appear in the images at resonance frequency due to
axial-radial coupling.}
\end{figure}
\begin{figure}[tbph]
\includegraphics[width=8.5 cm,bb=84pt 633pt 505pt 772pt]{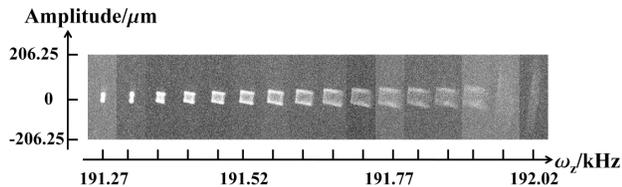}
\caption{The coupled oscillation in the time-average images for the positive
scan. Comparing to Fig. 3, the scan with smaller step (0.01 kHz) presents
clearer pictures for the rectangle trajectories from appearance to
disappearance, where sixteen images are selected in series with the
frequency step 0.05 kHz.}
\end{figure}

We measure the ion oscillation by taking time-averaged images from the
EMCCD. In Fig. 3, seventeen such images for different drive frequencies are
presented for positive and negative scans, respectively. For the ion
originally oscillating in the $z$-axis, we slowly scan the drive frequency $%
\omega_{z}$ across the harmonic resonance at $\omega_{0z}$. When the
detuning $\sigma =\omega_{z}-\omega_{0z}$ approaches zero, the rectangle
trajectory appears, implying a coupled motion between $x$- and $z$-axes due
to axial-radial coupling (explained later). For a more clarified
observation, we scan with smaller steps around the regime $\sigma=0$, as
shown in Fig. 4 which gives us an accurate range from the appearance of the
rectangle to the disappearance.

\section{theoretical model}

To understand the observation above, we have to consider the multipole
potential in the SET, which is given by \cite{Littich}
\begin{equation}
\phi _{i}\left( x,y,z\right) =\sum_{j=1}^{\infty }g_{ij}M_{j}Y_{j}\left(
x,y,z\right) ,
\end{equation}
where the subscript $i$ labels different electrodes and the subscript $j$ is
for the spherical harmonics $Y_{j}\left( x,y,z\right) $. Both $Y_{j}\left(
x,y,z\right) $ and the related parameters $M_{j}$ are defined in \cite%
{Littich}. $g_{ij}$ is the weight factor for different electrodes. In this
treatment, the initial equilibrium position of the single trapped ion is
defined as the origin of the coordinates. The five-wire SET generally
consists of quadrupole and hexapole potentials \cite{Blakestad}. Considering
the defect in our SET, we also involve octopole potential in our treatment.
Following the definition in \cite{Littich}, we have the subscripts $%
j=5,\cdots ,25$ where $j$ from 5 to 9, from 10 to 16 and from 17 to 25
correspond, respectively, to the quadrupole, hexapole and octopole
potentials. For different potentials $\Psi _{i}=V_{i}+U_{i}\cos (\Omega t)$
applied, respectively, to N electrodes, where $V_{i}$ is the dc voltage on
the electrode $i$ and $U_{i}\cos (\Omega t)$ represents the rf voltage $%
U_{i} $ on the electrode $i$ driven at frequency $\Omega $, we rewrite Eq.
(1) for the dc potential $\Phi _{dc}$ and the rf potential $\Phi _{rf}$ as
\begin{equation}
\begin{array}{c}
\Phi _{dc}\left( x,y,z\right) =\sum_{j=5}^{25}V_{j}^{\ast }M_{j}Y_{j}\left(
x,y,z\right) , \\
\Phi _{rf}\left( x,y,z\right) =\sum_{j=5}^{25}U_{j}^{\ast }\cos (\Omega
t)M_{j}Y_{j}\left( x,y,z\right),%
\end{array}%
\end{equation}%
where we have used $U_{j}^{\ast }=\sum_{i=1}^{N}U_{i}g_{ij}$ and $%
V_{j}^{\ast }=\sum_{i=1}^{N}V_{i}g_{ij}$.

Moreover, we have the 1D motional equation for the trapped ion \cite{Doroudi}%
,
\begin{eqnarray}
&&\frac{d^{2}\xi }{dt^{2}}+2\mu \frac{d\xi }{dt}+\frac{e}{m}\frac{\partial
\Phi _{dc}}{\partial \xi }+\frac{e^{2}}{2m^{2}}\frac{\partial }{\partial \xi
}\left( \left\langle \left\vert \int \frac{\partial \Phi _{rf}}{\partial \xi
}dt\right\vert ^{2}\right\rangle \right)   \notag \\
&=&k_{\xi }\cos (\omega _{\xi }t),
\end{eqnarray}%
with $\xi =x,y$ and $z$. Combining Eq. (2) with Eq. (3), we obtain the
equation of motion in the $z$ direction,
\begin{equation}
\begin{array}{c}
\frac{d^{2}z}{dt^{2}}+2\mu \frac{dz}{dt}+\omega _{0z}^{2}z+\alpha
_{2}z^{2}+\alpha _{3}z^{3}+\alpha _{21}z^{2}y+\alpha _{22}z^{2}x \\
+\alpha _{4}zy^{2}+\alpha _{5}zx^{2}+\alpha _{6}zxy+\alpha _{7}zy+\alpha
_{8}zx=k_{z}\cos (\omega _{z}t),%
\end{array}%
\end{equation}%
where $x$, $y$ and $z$ represent, respectively, the displacement of the ion
from the equilibrium position in the three dimensions, $\mu $ is the linear
damping parameter originated from the recoil due to photon absorption, $k_{z}
$ is driving amplitude. The detailed expressions of the nonlinear
coefficients $\alpha _{i}$ $(i=2,21,22,3,4,5,6,7,8)$ can be found in
Appendix I. Compared to the Duffing oscillator in \cite{aker}, Eq. (4) is a
complicated Duffing oscillator, containing additional coupled-motion terms.

Using the method of multiple scales \cite{S.S2001}, we obtain the
steady-state solution to Eq. (4) as
\begin{eqnarray}
\sigma  &=&\frac{3a^{2}}{8\omega _{0z}}(\alpha _{3}+\frac{-10\alpha _{2}^{2}%
}{9\omega _{0z}^{2}}+\frac{-2\alpha _{8}\alpha _{2x}c^{2}}{3\omega
_{0x}^{2}a^{2}}+\frac{-2\alpha _{7}\alpha _{2y}b^{2}}{3\omega _{0y}^{2}a^{2}}
\notag \\
&&+\frac{2\alpha _{4}b^{2}}{3a^{2}}+\frac{2\alpha _{5}c^{2}}{3a^{2}})\pm
\sqrt{\frac{k_{z}^{2}}{4\omega _{0z}^{2}a^{2}}-\mu ^{2}},
\end{eqnarray}%
where the nonlinear coefficients $\alpha _{2x}$ and $\alpha _{2y}$ are
relevant to the coupled motion along $x$- and $y$-axes. $c$, $b$ and $a$ are
the response amplitudes, respectively, in $x$, $y$ and $z$ directions. $%
\omega _{0x}/2\pi $ and $\omega _{0y}/2\pi $ represent the harmonic
frequencies in $x$- and $y$-axes. For more clarification, we define a
parameter $\alpha _{total}$ as
\begin{equation}
\alpha _{total}=\alpha _{3}+\Delta \alpha _{2}+\Delta \alpha ,
\end{equation}%
where $\alpha _{3}$ originates from the cubic nonlinearity, $\Delta \alpha
_{2}=\frac{-10\alpha _{2}^{2}}{9\omega _{0z}^{2}}$ represents the nonlinear
coefficient that comes from quadric nonlinearity, and $\Delta \alpha =\frac{%
-2\alpha _{8}\alpha _{2x}c^{2}}{3\omega _{0x}^{2}a^{2}}+\frac{-2\alpha
_{7}\alpha _{2y}b^{2}}{3\omega _{0y}^{2}a^{2}}+\frac{2\alpha _{4}b^{2}}{%
3a^{2}}+\frac{2\alpha _{5}c^{2}}{3a^{2}}$ corresponds to the nonlinear
dispersion relevant to the coupled motion. Substituting $\alpha _{total}$
into Eq. (5), we obtain
\begin{equation}
\sigma =\frac{3\alpha _{total}}{8\omega _{0z}}a^{2}\pm \sqrt{\frac{k_{z}^{2}%
}{4\omega _{0z}^{2}a^{2}}-\mu ^{2}}.
\end{equation}

\section{discussion about the nonlinearity and coupling}

In our home-built SET, since the harmonic frequency in y-axis is much bigger
than in other axes, the ion is confined very tightly in y direction, which
leads to a reasonable assumption $b/a\ll $1. As a result, the coupled term $%
\Delta \alpha $ is reduced to $\Delta \alpha =\chi c^{2}/a^{2}$ with $\chi =%
\frac{-2\alpha _{8}\alpha _{2x}}{3\omega _{0x}^{2}}+\frac{2\alpha _{5}}{3}$.
Moreover, $\alpha _{3}$ and $\Delta \alpha _{2}$ in Eq. (6) are nothing to
do with the coupled motion and their sum $\alpha _{3}+\Delta \alpha _{2}$
can be measured experimentally by
\begin{equation}
a_{m}=\sqrt{\frac{8\omega _{0z}\sigma _{m}}{3(\alpha _{3}+\Delta \alpha _{2})%
}},
\end{equation}%
with the maximal amplitude $a_{m}$ and the maximal detuning $\sigma _{m}$ in
the non-coupling case. As a result, Eq. (7) is reduced to a steady-state
solution to the amplitude of the response $a$ with respect to the drive
detuning $\sigma $ for the known driving force amplitude $k_{z}$,
\begin{equation}
\sigma =\frac{3a^{2}}{8\omega _{0z}}(\alpha _{3}+\Delta \alpha _{2}+\chi
\frac{c^{2}}{a^{2}})\pm \sqrt{\frac{k_{z}^{2}}{4\omega _{0z}^{2}a^{2}}-\mu
^{2}}.
\end{equation}


\begin{figure}[tbph]
\includegraphics[width=8.5 cm,bb=57pt 198pt 459pt 644pt]{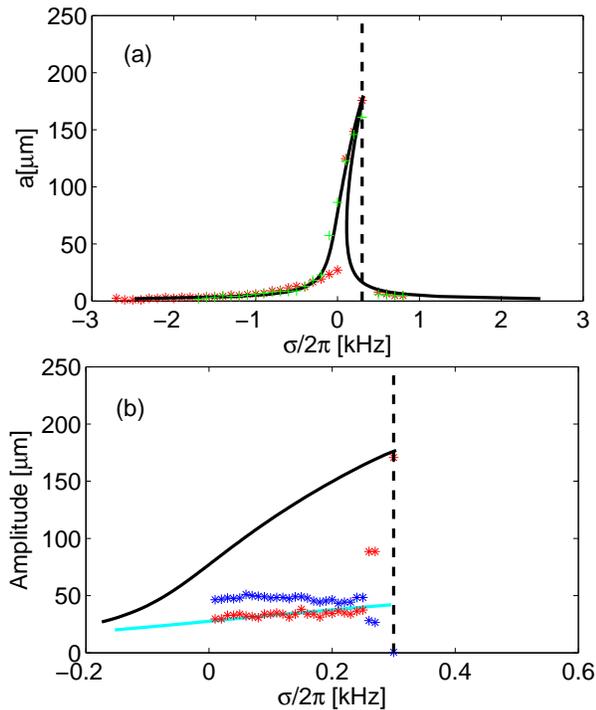}
\caption{The measured and calculated amplitudes with respect to the driving
detuning, where the black dashed lines indicate the detunings for maximum
amplitudes of oscillations. (a) The measured values correspond to the
situation in Fig. (3). Calculation (black solid curve) by Eq. (9) without
the coupling term fits most of the measured values with positive scan (red
stars) and negative scan (green crosses). The scan step is $0.1$ kHz. (b)
The measured values correspond to the situation in Fig. (4). Calculation
(light blue solid line) by Eq. (9) with the coupling term fits the measured
values with positive slow scan (red stars) around zero detuning. The scan
step is $0.01$ kHz, and the blue stars are for the measured values of the
vibrational amplitude $c$ in $x$-axis. As a comparison, the calculation
without the coupling term is plotted (the black solid curve).}
\end{figure}

In our experiment, $\omega _{0z}$ is measured via ion response in the linear
regime, $k_{z}=0.075\times 10^{6}$ Hz$^{2}$m is obtained by observing the
ion displacement versus the middle electrode voltage \cite{house}, $\alpha
_{3}+\Delta \alpha _{2}=0.1959\times 10^{18}$ Hz $^{2}$/m$^{2}$ is measured
using the observed dependence of $a_{m}$ on the maximal detuning $\sigma
_{m} $. We evaluate $\mu =177.1$ Hz using the relation $a_{m}=k_{z}/\left(
2\mu \omega _{0z}\right) $. The comparison in the non-coupling case between
the measured and calculated values of $a$ and $\sigma $ is made in Fig.
5(a), where Eq. (9) without the coupling term $\Delta \alpha $ (the black
solid curve) can fit most experimental values for both the positive and
negative scans (red stars and green crosses, respectively). In this
situation, the vibrational amplitude $c$ in $x$-axis is negligible. Some
experimental values around $\sigma =0$, which are not fitted well by the
solid curve, are actually relevant to the case of coupled motion. To be more
clarified, we scan the region around $\sigma =0$ with smaller step than in
Fig. 5(a). The fitting by considering the coupling term $\Delta \alpha $ in
our calculation can fully cover the measured data, as shown in Fig. 5(b). In
such a case, we find that the vibrational amplitude $c$ in $x$-axis is
visible, which is excited by the energy transfer from $z$-axis due to
motional coupling. This energy transferred from $z$-axis to $x$-axis is
nearly constant in the adiabatic operation so that we obtain $\chi \frac{%
c^{2}}{a^{2}}\approx 4.5\times 10^{18}$ Hz$^{2}$/m$^{2}$. Fig. 5(b) also
shows that the motional coupling stops when $\sigma/2\pi$ approaches $0.25$ kHz.
We see that $a$ goes up to a maximum with $c$ dropping to zero, implying
that the system returns to the non-coupling case. Therefore, the vibrational
trajectories imaged in Figs. 3 and 4 can be fully understood by the
complicated Duffing oscillator with and without the term for coupled motion.

\begin{figure}[tbph]
\includegraphics[width=8.5 cm]{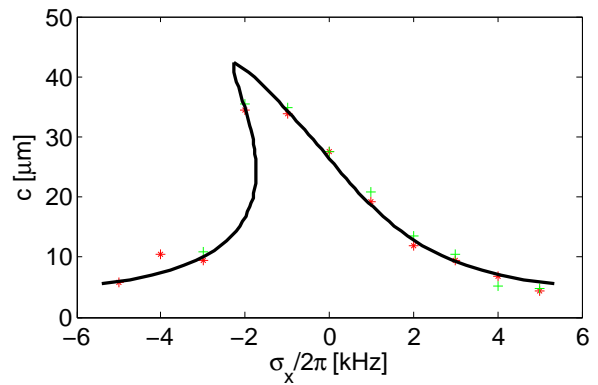}
\caption{The measured and calculated amplitudes in $x$-axis with respect to
the driving detuning, where the scan step is 1 kHz and the measurement is
made in the no-coupling case. Calculation (black solid curve) fits most of
the measured values with positive scan (red stars) and negative scan (green
crosses). }
\end{figure}

Moreover, we also checked nonlinear effects in different directions in our
SET by applying the drive on the $x$-axis and repeating the experimental
steps as above for $z$-axis. It is physically evident that the behavior can
be described by a slight modification of Eq. (4) by exchanging $z$ and $x$,
and replacing $k_{z}$ and $\omega_{z}$ by $k_{x}$ and $\omega_{x}$. As an
example, we only present the non-coupling case in Fig. 6, where the measured
data is fitted well by the steady solution to the complicated Duffing
oscillator with different nonlinear coefficients and different damping rates
from in Fig. 5(a). In comparison to the shape of the curve in Fig. 5(a), the
oscillation in such a case reaches the maximum amplitude before $\sigma =0$,
i.e, the red detuning, corresponding to $\alpha_{3}+\Delta \alpha_{2}<0$ in
Eq. (8). This implies negative coefficients of quadratic and cubic terms in
Eq. (4) originated from the different asymmetry from in $z $-axis.

\section{conclusion}

In conclusion, we have experimentally investigated the complicated
oscillations in our home-built SET, which are related to the high-order
multipole potentials. Both the coupling and non-coupling cases, as well as
the driving along different axes, are studied. Our observation can be fully
understood by the nonlinear effects and the motional coupling from the
solution of a complicated Duffing oscillator.

In comparison to the relevant study on a single ion oscillating in the
linear trap \cite{aker}, our home-made SET owns higher-order multipole
potentials, which cause more fruitful nonlinear effects and even the
motional coupling between different directions. Although there are also
axial-radial couplings observed in the ion trap, e.g., with ion cloud in \cite%
{cloud}, such a motional coupling is much more evident in the SET, which, in
addition to the couplings regarding $zx$ and $xz^{2}$, is also reflected in
the $zx^{2}$ term in Eq. (4) with the coefficient $\alpha _{5}$. According
to our calculation, the motional coupling in our observation is mainly
influenced by the coefficients $\alpha _{5}$ and $\alpha _{8}$, which
implies the combined action from the quadrupole, hexapole and octopole
potentials. This is the reason that the rectangle trajectories have never
been observed previously in linear ion traps. Moreover, recent investigation
of the phonon laser based on the nonlinear oscillation of the trapped ion
demonstrated the analogy to the Fabry-Perot laser with 100\% reflecting
mirrors \cite{vahala}. In contrast, the SET under our study seems an
asymmetry Fabry-Perot cavity, which may yield two split beams of the phonon
laser in perpendicular axes. Coherent transfer between the two split phonon beams
would be useful in fundamental physics and practical application. Further work
in this aspect is underway.

With the trapped ions cooled down to the motional ground state, we may have
an excellent platform to demonstrate nonlinear behavior in a fully quantum
mechanical regime and also carry out quantum logic gate operations.
Therefore our work presents a way to exploring complicated nonlinearity
using experimentally available techniques and it is also useful for
suppressing detrimental effects from the nonlinearity in quantum information
processing using trapped ions.

\section*{Acknowledgments}

This work is supported by National Fundamental Research Program of China
under Grant No. 2012CB922102, and by National Natural Science Foundation of
China under Grants No. 11274352 and No. 11104325.

\begin{widetext}
\appendix

\section*{Appendix I~~~~ The nonlinear coefficients in Eq. (4)}

Our calculation is based on Eq. (4), in which the nonlinear coefficients
originate from the high-order multipole potentials. By setting $e$ and $m$
to be the electric quantity and the mass of single calcium ion, we have the
nonlinear coefficients as
\begin{equation}
\alpha _{3}=\frac{36e^{2}r_{0}^{2}\left( U_{13}^{\ast }\right) ^{2}M_{13}^{2}%
}{m^{2}r_{0}^{8}\Omega ^{2}}+\frac{140eM_{21}\left[ mr_{0}^{4}(V_{21}^{\ast
})\Omega ^{2}+8er_{0}^{2}\left( U_{21}^{\ast }\right) \left( U_{7}^{\ast
}\right) M_{7}\right] }{m^{2}r_{0}^{8}\Omega ^{2}},
\end{equation}%
\begin{equation}
\alpha _{2}=\frac{6er_{0}M_{13}\left[ mr_{0}^{4}(V_{13}^{\ast })\Omega
^{2}+6er_{0}^{2}\left( U_{7}^{\ast }\right) \left( U_{13}^{\ast }\right)
M_{7}\right] }{m^{2}r_{0}^{8}\Omega ^{2}},
\end{equation}%
\begin{eqnarray}
\alpha _{21} &=&\frac{3e\left[ 70er_{0}^{2}M_{6}M_{21}(U_{6}^{\ast
})(U_{21}^{\ast })+42er_{0}^{2}M_{7}M_{20}(U_{7}^{\ast })(U_{20}^{\ast })%
\right] }{m^{2}r_{0}^{8}\Omega ^{2}}  \notag \\
&&+\frac{3e\left[ 24er_{0}^{2}M_{12}M_{13}(U_{12}^{\ast })(U_{13}^{\ast
})+7mr_{0}^{4}\Omega ^{2}M_{20}(V_{20}^{\ast })\right] }{m^{2}r_{0}^{8}%
\Omega ^{2}},
\end{eqnarray}%
\begin{eqnarray}
\alpha _{22} &=&\frac{3e\left[ 70er_{0}^{2}M_{8}M_{21}(U_{8}^{\ast
})(U_{21}^{\ast })+42er_{0}^{2}M_{7}M_{22}(U_{7}^{\ast })(U_{22}^{\ast })%
\right] }{m^{2}r_{0}^{8}\Omega ^{2}}  \notag \\
&&+\frac{3e\left[ 24er_{0}^{2}M_{13}M_{14}(U_{13}^{\ast })(U_{14}^{\ast
})+7mr_{0}^{4}\Omega ^{2}M_{22}(V_{22}^{\ast })\right] }{m^{2}r_{0}^{8}%
\Omega ^{2}},
\end{eqnarray}%
\begin{eqnarray}
\alpha _{4} &=&\frac{e\left[ 32er_{0}^{2}\left( U_{12}^{\ast }\right)
^{2}M_{12}^{2}-18er_{0}^{2}\left( U_{13}^{\ast }\right)
^{2}M_{13}^{2}-6er_{0}^{2}\left( U_{13}^{\ast }\right) \left( U_{15}^{\ast
}\right) M_{13}M_{15}+21er_{0}^{2}\left( U_{6}^{\ast }\right) \left(
U_{20}^{\ast }\right) M_{6}M_{20}\right] }{m^{2}r_{0}^{8}\Omega ^{2}}  \notag
\\
&&+\frac{e\left[ -60mr_{0}^{4}(V_{21}^{\ast })\Omega
^{2}M_{21}-14mr_{0}^{4}(V_{23}^{\ast })\Omega
^{2}M_{23}-240er_{0}^{2}(U_{7}^{\ast })\left( U_{21}^{\ast }\right)
M_{7}M_{21}-56er_{0}^{2}(U_{7}^{\ast })\left( U_{23}^{\ast }\right)
M_{7}M_{23}\right] }{m^{2}r_{0}^{8}\Omega ^{2}},
\end{eqnarray}

\begin{eqnarray}
\alpha _{5} &=&\frac{e\left[ 32er_{0}^{2}\left( U_{14}^{\ast }\right)
^{2}M_{14}^{2}-18er_{0}^{2}\left( U_{13}^{\ast }\right)
^{2}M_{13}^{2}+6er_{0}^{2}\left( U_{13}^{\ast }\right) \left( U_{15}^{\ast
}\right) M_{13}M_{15}+21er_{0}^{2}M_{8}M_{22}(U_{8}^{\ast })(U_{22}^{\ast })%
\right] }{m^{2}r_{0}^{8}\Omega ^{2}}  \notag \\
&&+\frac{e\left[ -60mr_{0}^{4}(V_{21}^{\ast })\Omega
^{2}M_{21}+14mr_{0}^{4}(V_{23}^{\ast })\Omega
^{2}M_{23}-240er_{0}^{2}(U_{7}^{\ast })\left( U_{21}^{\ast }\right)
M_{7}M_{21}+56er_{0}^{2}(U_{7}^{\ast })\left( U_{23}^{\ast }\right)
M_{7}M_{23}\right] }{m^{2}r_{0}^{8}\Omega ^{2}},
\end{eqnarray}%
\begin{eqnarray}
\alpha _{6} &=&\frac{e\left[ 6er_{0}^{2}\left( U_{13}^{\ast }\right) \left(
U_{11}^{\ast }\right) M_{13}M_{11}+14mr_{0}^{4}(V_{19}^{\ast })\Omega
^{2}M_{19}+21er_{0}^{2}\left( U_{6}^{\ast }\right) \left( U_{22}^{\ast
}\right) M_{6}M_{22}\right] }{m^{2}r_{0}^{8}\Omega ^{2}}  \notag \\
&&+\frac{e\left[ 64er_{0}^{2}(U_{12}^{\ast })\left( U_{14}^{\ast }\right)
M_{12}M_{14}+56er_{0}^{2}\left( U_{7}^{\ast }\right) \left( U_{19}^{\ast
}\right) M_{7}M_{19}+21er_{0}^{2}M_{8}M_{20}(U_{8}^{\ast })(U_{20}^{\ast })%
\right] }{m^{2}r_{0}^{8}\Omega ^{2}},
\end{eqnarray}%
\begin{equation}
\alpha _{7}=\frac{e\left[ 6er_{0}^{3}\left( U_{13}^{\ast }\right) \left(
U_{6}^{\ast }\right) M_{13}M_{6}+8mr_{0}^{5}(V_{12}^{\ast })\Omega
^{2}M_{12}+32er_{0}^{3}\left( U_{12}^{\ast }\right) \left( U_{7}^{\ast
}\right) M_{12}M_{7}\right] }{m^{2}r_{0}^{8}\Omega ^{2}},
\end{equation}%
\begin{equation}
\alpha _{8}=\frac{e}{m^{2}r_{0}^{8}\Omega ^{2}}\left[ 8mr_{0}^{5}(V_{14}^{%
\ast })\Omega ^{2}M_{14}+32er_{0}^{3}\left( U_{14}^{\ast }\right) \left(
U_{7}^{\ast }\right) M_{14}M_{7}+6er_{0}^{3}M_{8}M_{13}(U_{8}^{\ast
})(U_{13}^{\ast })\right] ,
\end{equation}%
with $\omega _{0z}^{2}=e\left[ 8er_{0}^{4}\left( U_{7}^{\ast }\right)
^{2}M_{7}^{2}+4mr_{0}^{6}(V_{7}^{\ast })\Omega ^{2}M_{7}\right]
/m^{2}r_{0}^{8}\Omega ^{2}$, and the scaling factor $r_{0}$ (See definition
in \cite{Littich}).

\appendix

\section*{Appendix II~~~~ Details of the steady-state solution Eq. (5)}

Eq. (5) is obtained by the standard steps of the multiple scale method.
Starting from Eq. (4), we assume that the driving frequency is a
perturbative expansion of harmonic oscillator frequency \cite{Nayfeh1985},
i.e., $\omega _{z}=\omega _{0z}+\epsilon ^{2}\sigma $, with a small
dimensionless parameter $\epsilon $. Following the method in \cite%
{Nayfeh1984}, we rewrite Eq. (4) by setting $z=\epsilon u$, $x=\epsilon p$, $%
y=\epsilon q$, the damping term as $2\epsilon ^{3}\mu \dot{u}$ and the
driving term as $\epsilon ^{3}k_{z}\cos (\omega _{z}t)$. Introducing a new
parameter $T_{i}=\epsilon ^{i}t$ $(i=0,1,2)$, we rewrite $u,p,q$ as,
\begin{equation}
\begin{array}{c}
u=\epsilon ^{0}u_{0}\left( T_{0},T_{1},T_{2}\right) +\epsilon
^{1}u_{1}\left( T_{0},T_{1},T_{2}\right) +\epsilon ^{2}u_{2}\left(
T_{0},T_{1},T_{2}\right) , \\
p=\epsilon ^{0}p_{0}\left( T_{0},T_{1},T_{2}\right) +\epsilon
^{1}p_{1}\left( T_{0},T_{1},T_{2}\right) +\epsilon ^{2}p_{2}\left(
T_{0},T_{1},T_{2}\right) , \\
q=\epsilon ^{0}q_{0}\left( T_{0},T_{1},T_{2}\right) +\epsilon
^{1}q_{1}\left( T_{0},T_{1},T_{2}\right) +\epsilon ^{2}q_{2}\left(
T_{0},T_{1},T_{2}\right) .%
\end{array}%
\end{equation}%
Then we compare the coefficients of $\epsilon ^{0}$, $\epsilon ^{1}$ and $%
\epsilon ^{2}$, which yields,
\begin{equation}
D_{0}^{2}u_{0}+\omega _{0z}^{2}u_{0}=0,
\end{equation}%
\begin{equation}
D_{0}^{2}u_{1}+\omega _{0z}^{2}u_{1}=-2D_{0}D_{1}u_{0}-\alpha
_{2}u_{0}^{2}-\alpha _{7}q_{0}u_{0}-\alpha _{8}p_{0}u_{0},
\end{equation}%
\begin{eqnarray}
D_{0}^{2}u_{2}+\omega _{0z}^{2}u_{2} &=&-[-k_{z}\cos \left( \omega
_{0z}T_{0}+\sigma T_{2}\right) +q_{0}^{2}\alpha _{4}u_{0}+p_{0}^{2}\alpha
_{5}u_{0}+p_{0}q_{0}\alpha _{6}u_{0}+2\mu D_{0}u_{0}  \notag \\
&&+D_{1}^{2}u_{0}+2D_{0}D_{2}u_{0}+q_{0}\alpha _{21}u_{0}^{2}+p_{0}\alpha
_{22}u_{0}^{2}+\alpha _{3}u_{0}^{3}+2D_{0}D_{1}u_{1}  \notag \\
&&+\alpha _{7}q_{0}u_{1}+\alpha _{8}p_{0}u_{1}+2\alpha _{2}u_{0}u_{1}+\alpha
_{7}q_{1}u_{0}+\alpha _{8}p_{1}u_{0}],
\end{eqnarray}%
where $D_{i}=\partial /\partial T_{i}$, $\left( i=0,1,2\right) $.

We may solve $u_{0}=A\left( T_{2}\right) \exp (i\omega _{0z}T_{0})+\bar{A}%
\left( T_{2}\right) \exp (-i\omega _{0z}T_{0})$ and $u_{1}=\alpha
_{2}[A^{2}\exp (2i\omega _{0z}T_{0})-6A\bar{A}+\bar{A}^{2}\exp (-2i\omega
_{0z}T_{0})]/3\omega _{0z}^{2}$ from Eqs. (20) and (21) by eliminating the
secular term. Similarly, from equations of the oscillations in x-axis and
y-axis, we may solve the variables $p_{0}$, $q_{0}$, $p_{1}$ and $q_{1}$ as
\begin{equation}
p_{0}=C\left( T_{2}\right) \exp (i\omega _{0x}T_{0})+\bar{C}\left(
T_{2}\right) \exp (-i\omega _{0x}T_{0}),
\end{equation}%
\begin{equation}
q_{0}=B\left( T_{2}\right) \exp (i\omega _{0y}T_{0})+\bar{B}\left(
T_{2}\right) \exp (-i\omega _{0y}T_{0}),
\end{equation}%
\begin{equation}
p_{1}=\alpha _{2x}[C^{2}\exp (2i\omega _{0x}T_{0})-6C\bar{C}+\bar{C}^{2}\exp
(-2i\omega _{0x}T_{0})]/(3\omega _{0x}^{2}),
\end{equation}%
\begin{equation}
q_{1}=\alpha _{2y}[B^{2}\exp (2i\omega _{0y}T_{0})-6B\bar{B}+\bar{B}^{2}\exp
(-2i\omega _{0y}T_{0})]/(3\omega _{0y}^{2}),
\end{equation}%
where $\alpha _{2x}$ and $\alpha _{2y}$ correspond to the quadric
nonlinearity of the ion motion equation in $x$ and $y$ directions,
respectively. $\omega _{0x}/2\pi $ and $\omega _{0y}/2\pi $ represent the
harmonic frequencies in x direction and y direction. $A=\frac{1}{2}a\exp
(i\beta )$, $B=\frac{1}{2}b\exp (i\varsigma )$ and $C=\frac{1}{2}c\exp
(i\eta )$, where $a,b,c,\beta ,\varsigma $ and $\eta $ are real functions of
$T_{2}$, $\beta ,\varsigma $ and $\eta $ represent the phases of different
dimensions. $\bar{A}$, $\bar{B}$ and $\bar{C}$ are conjugate terms of A, B
and C. Substituting $u_{0}$ and $u_{1}$ into Eq. (22), we obtain an equation
regarding the secular term, from which, in combination of Eqs. (23-26) with
the expressions of $A$, $B$ and $C$, we obtain
\begin{equation}
-\frac{1}{2}k_{z}\sin \gamma +a\mu \omega _{0z}+\omega _{0z}\frac{da}{dT_{2}}%
=0,
\end{equation}%
\begin{equation}
\frac{1}{2}k_{z}\cos \gamma -\frac{3\alpha _{3}a^{3}}{8}-\frac{1}{4}\alpha
_{4}ab^{2}-\frac{1}{4}\alpha _{5}ac^{2}+\frac{5\alpha _{2}^{2}a^{3}}{%
12\omega _{0z}^{2}}+\frac{\alpha _{8}\alpha _{2x}ac^{2}}{4\omega _{0x}^{2}}+%
\frac{\alpha _{7}\alpha _{2y}ab^{2}}{4\omega _{0y}^{2}}+a\omega _{0z}\left(
\sigma -\frac{d\gamma }{dT_{2}}\right) =0,
\end{equation}%
with $\gamma =\sigma T_{2}-\beta $. We assume the steady-state motion
corresponding to $\frac{d\gamma }{dT_{2}}=\frac{d\gamma }{dT_{2}}=0$. So we
have
\begin{equation}
\left( \frac{3\alpha _{3}}{8}a^{3}-\frac{5\alpha _{2}^{2}}{12\omega _{0z}^{2}%
}a^{3}+\frac{\alpha _{4}}{4}ab^{2}+\frac{\alpha _{5}}{4}ac^{2}-\frac{\alpha
_{7}\alpha _{2y}}{4\omega _{0y}^{2}}ab^{2}-\frac{\alpha _{8}\alpha _{2z}}{%
4\omega _{0x}^{2}}ac^{2}-a\omega _{0z}\sigma \right) ^{2}+\left( a\omega
_{0z}\mu \right) ^{2}=\frac{1}{4}k_{z}^{2}.
\end{equation}%
which is actually Eq. (5).
\end{widetext}


\begin{thebibliography}{99}
\bibitem{Nayfeh1979} A. H. Nayfeh and D. T. Mook, \textit{Nonlinear
Oscillations} (Wiley-Interscience, New York, 1979).

\bibitem{M.I. Dykman} M. I. Dykman and M. A. Krivoglaz, \textit{Soviet
Scientific Reviews} Volume 5, 265 (Harwood Academic, 1984).

\bibitem{L. D. Landau} L. D. Landau and E. M. Lifshitz. \textit{"Mechanics"}
(Pergamon, New York, 3rd edition, 1976).

\bibitem{Nayfeh1981} A. H. Nayfeh. \textit{Introduction to Perturbation
Techniques} (Wiley, New York, 1981).

\bibitem{V. I. Arnold} V. I. Arnold. \textit{Geometrical methods in the
theroy of ordinaty differential equations}, volume 250 of \textit{%
Grundlehren der mathematischen Wissenschaften} (Springer-Verlag, New York,
2nd edition, 1988).

\bibitem{S. H. Strogatz} S. H. Strogatz. \textit{Nonlinear Dynamics and
Chaos: with applications to physics, biology, chemistry, and engineering}
(Perseus Books, 1994).

\bibitem{H. B. Chan} H. B. Chan, M.I. Dykman, and C. Stambaugh, Phys. Rev.
Lett. \textbf{100}, 130602 (2008).

\bibitem{M. I. Dylman2004} M. I. Dykman, B. Golding, and D. Ryvkine, Phys.
Rev. Lett. \textbf{92}, 080602 (2004).

\bibitem{B. Yurke} B. Yurke and E. Buks, J. Lightwave Tech. \textbf{24},
5054 (2006).

\bibitem{E. Buks} E. Buks and B. Yurke, Phys. Rev. A \textbf{73}, 23815
(2006).

\bibitem{blatt} D. Leibfried, R. Blatt, C. Monroe and D. Wineland, Rev. Mod.
Phys. \textbf{75}, 281 (2003).

\bibitem{W. Paul} W. Paul, Rev. Mod. Phys. \textbf{62}, 531 (1990).

\bibitem{aker} N. Akerman, S. Kotler, Y. Glickman, Y. Dallal, A. Keselman,
and R. Ozeri, Phys. Rev. A \textbf{82}, 061402(R) (2010).

\bibitem{mak} A. A. Makarov, Anal. Chem. \textbf{68}, 4257 (1996).

\bibitem{dra} A. Drakoudis, M. S{\"o}llner and G. Werth, Int. J. Mass.
Spectrom. \textbf{252}, 61 (2006).

\bibitem{vahala} K. Vahala, M. Herrmann, S. Kn{\"u}nz, V. Batteiger, G.
Saathoff, T. W. H{\"u}nsch and Th. Udem, Nat. Phys. \textbf{5}, 682 (2009);
S. Kn{\"u}nz, M. Herrmann, V. Batteiger, G. Saathoff, T. W. H{\"u}nsch, K.
Vahala and Th. Udem, Phys. Rev. Lett. \textbf{105}, 013004 (2010).

\bibitem{wineland} D. Kielpinksi, C. Monroe and D. J. Wineland, Nature
(London) \textbf{417}, 709 (2002).

\bibitem{wesen} J. H. Wesenberg, Phys. Rev. A \textbf{78}, 063410 (2008).

\bibitem{house} M. G. House, Phys. Rev. A \textbf{78}, 033402 (2008).

\bibitem{Blakestad} R. Bradford Blakestad, \textit{Transport of Trapped-Ion
Qubits within a Scalable Quantum Processor} [D] (California Institute of
Technology, 2002).

\bibitem{chip} L. Chen, W. Wan, Y. Xie, H.-Y. Wu, F. Zhou and M. Feng, Chin.
Phys. Lett. \textbf{30}, 013702 (2013).

\bibitem{njp} D. T. C. Allcock, J. A. Sherman, D. N. Stacey, A. H. Burrell,
M. J. Curtis, G. Imreh, N. M. Linke, D. J. Szwer, S. C. Webster, A. M.
Steane and D. M. Lucas, New J. Phys. \textbf{12}, 053026 (2010).

\bibitem{Littich} Gebhard Littich, \textit{Electrostatic Control and
Transport of Ions on a Planar Trap for Quantum Information Processing} [D]
(ETH Z\"{u}rich and University of California, Berkeley, 2011).

\bibitem{Doroudi} A. Doroudi, Phys. Rev. E \textbf{80}, 056603 (2009).

\bibitem{S.S2001} S. Sevugarajan and A. G. Menon, Int. J. Mass Spectrom.
\textbf{209}, 209 (2001).

\bibitem{cloud} M. Vedel, J. Rocher, M. Knoop and F. Vedel, Appl. Phys. B
\textbf{66}, 191 (1998).

\bibitem{Nayfeh1985} A. H. Nayfeh, \textit{Problems in Perturbation}
(Wiley-Interscience, New York, 1985).

\bibitem{Nayfeh1984} A. H. Nayfeh, J. Sound Vib. \textbf{92}, 363 (1984).
\end{thebibliography}
\end{document}